\documentclass[11pt,preprint]{aastex}

\newcommand\be{\begin{equation}}
\newcommand\en{\end{equation}}

\shorttitle{Coronet cluster}
\shortauthors{Sicilia-Aguilar et al.}

\begin{document}

\title{Protostars and stars in the Coronet cluster: Age, evolution, and cluster structure\thanks{Based on observations collected at the European Southern Observatory, Paranal, Chile (Proposal IDs: 081.C-0204(A), 083.C-0079(A), 083.C-0079(B))}
}
\author{Aurora Sicilia-Aguilar\altaffilmark{1}, Thomas Henning\altaffilmark{1}, Jouni Kainulainen\altaffilmark{1}, }
\author{Veronica Roccatagliata\altaffilmark{2}}
\altaffiltext{1}{Max-Planck-Institut f\"{u}r Astronomie, K\"{o}nigstuhl 17, 69117 Heidelberg, Germany}
\altaffiltext{2}{Space Telescope Science Institute, Baltimore, MD 21218, USA}

\email{sicilia@mpia.de}

\begin{abstract}
We present new optical spectroscopy with FLAMES/VLT, near-IR imaging with HAWK-I/VLT, 
and 870$\mu$m mapping with APEX/LABOCA
of the Coronet cluster. The optical data allow to estimate spectral types,
extinction and the presence of accretion in 6 more M-type members, in addition to the
12 that we had previously studied. The submillimeter
maps and near-IR data reveal the presence of nebular structures and high extinction
regions, which are in some cases associated to known IR, optical, and X-ray sources.
Most star formation is associated to two elongated structures crossing in the central 
part of the cluster. Placing all the 18 objects with known spectral types and extinction
in the HR diagram suggests that the cluster is younger than previously
thought ($<$2 Myr, and probably $\sim$0.5-1 Myr). The new age estimate is
in agreement with the evolutionary status of the various protostars
in the region and with its compactness ($<$1.3 pc across), but results in a
conflict with the low disk and accretion fraction (only 50-65\% of low-mass stars appear to
have protoplanetary disks, and most transitional and homologously depleted disks are
consistent with no accretion) and with the evolutionary features observed in
the mid-IR spectra and spectral energy distributions of the disks.

\end{abstract}


\section{Introduction \label{intro}}

The Coronet Cluster (also known as CrA) is an obscured star-forming region, 
located at 170 pc (Knude \& Hog 1998), associated to the Herbig Ae star R~CrA (Taylor \& Storey 1984) 
and a dense molecular core (Loren 1979). Submillimeter, millimeter, 
IR, and optical surveys (Henning et al. 1994; 
de Muizon et al. 1980; Wilking et al. 1985; Marraco \& Rydgren 1981) found a few HAeBe 
and T Tauri stars (TTS), suggesting a modest young stellar population of $\sim$1 Myr in age. 
X-ray surveys using Einstein, 
XMM-Newton, ROSAT and Chandra  (Walter 1986, 1997; Neuh\"{a}user et al. 2000; 
Hamaguchi et al. 2005; Garmire \& Garmire 2003; Forbrich \& Preibisch 2007),
less affected by extinction, identified 
more than 70 sources consistent with young stars, many of which are deeply embedded, all
located within a region a bit smaller than 1 pc in diameter. 
The presence of very embedded sources, including Class 0 objects, suggested a very
young region (Henning et al. 1994;
Chen et al. 1997; Chini et al. 2003; Groppi et al. 2004,2007; Nisini et al. 2005; Nutter et al. 2005).
Optical studies also revealed an important population of low-mass stars,
characterized by strong H$\alpha$ emission (L\'{o}pez-Mart\'{\i} et al. 2005).
The variety of sources makes the small Coronet cluster a very
special place: containing Class 0, I, II, and III objects, it seems to be
a relatively quiescent cluster, part of a larger molecular cloud where star
formation has occurred only in the densest parts.

Spitzer IRAC/MIPS data and IRS spectra
revealed more than 100 low and very low-mass sources, mostly Class I, 
Class II, and Class III objects, some of them highly extincted, 
constituting a very young and 
relatively rich cluster (Sicilia-Aguilar et al. 2008, from now on
SA08; Currie \& Sicilia-Aguilar 2011). Remarkably, the IRS spectra of
the disks around M-type cluster members (which are the bulk of the cluster members) 
reveal features that are usually accepted as signs of disk evolution:
lack of silicate emission, spectral energy distributions (SED) 
consistent with flattened/settled disks, very low or absent near-IR 
excesses characteristic of optically thin inner disk/inner holes (SA08). In
addition, the estimated disk fraction is lower than expected for a 
young, embedded cluster, with only 2/3 of the stars being surrounded
by protoplanetary disks. Moreover, recent membership searches 
(L\'{o}pez-Mart\'{\i} et al. 2010) have found additional diskless
low-mass members, which would reduce the disk fraction to about 50\%,
a number more consistent with 2-4 Myr old regions than with a
1-2 Myr old cluster (Sicilia-Aguilar et al. 2006; Hern\'{a}ndez et al. 2007; Fedele et al. 2010).

Here we present new FLAMES/VLT optical spectroscopy, deep IR imaging
with HAWK-I/VLT, and a 870$\mu$m map taken with APEX/LABOCA of the
Coronet cluster. The data provide spectral types and accretion 
constrains for most of the remaining low-mass cluster members with 
moderate extinction, as well as a detailed image of the structures and
substructures in the region, with associated Class 0/I embedded sources.
The datasets and reduction techniques are discussed in Section \ref{data}.
Spectral types, accretion, extinction maps, and the
cluster structure and embedded members are presented in Section \ref{analysis}.
In Section \ref{age} we discuss the age of the Coronet cluster based on
the identification of Class 0/I/II/III sources and cluster
structure, and finally our conclusions are presented in Section \ref{conclu}.

\section{Observations and data reduction \label{data}}

\subsection{FLAMES/VLT optical spectroscopy \label{flames}}

A large number of members and potential members had been 
already observed with FLAMES on the UT2/VLT (see SA08).
The remaining candidates were done as part of the ESO program
083.C-0079, following a similar scheme to our previous
observations. The data were taken in 5 nights (2009 May 24,
and 2009 June 3, 17, 18, 19), using FLAMES  
together with the multifiber spectrograph GIRAFFE and the 
MEDUSA fibers, which allow to take up to 130 spectra, covering
both objects and sky positions. We observed one field centered
near the cluster center with three different, intermediate-resolution
(R$\sim$5600-8600), gratings (L682.2, L773.4, L881.7). The
resulting spectra cover thus a large wavelength range from 6440\AA\
to 9400\AA, which includes the main accretion features (H$\alpha$,
Ca II), the Lithium I 6708\AA\ absorption line (indicative of youth),
and a large number of photospheric features and absorption bands, required for
a spectral type classification of M-type stars. We obtained 3$\times$2700~s 
exposures with each filter, which ensures good removal of cosmic rays.

The target list included the potential members detected via
X-ray (Garmire \& Garmire 2003) that had not been observed
in our previous run, together with a large number of sky positions
(more than 70 in each one of the filters)
to ensure a good background subtraction in this region with 
variable nebular emission. In addition, we also assigned fibers
to potential Herbig Haro (HH) objects in the cloud (Wang et al. 2004)
and to the objects that, having being detected in our previous
FLAMES observations, presented low S/N which made difficult 
a spectral type classification (especially, the disked stars
G-85 and G-87). In total, 12 sources were detected (see Table \ref{lines-tab}).
The lack of detections among the remaining X-ray candidates
confirms their status as extragalactic or highly embedded sources, 
as presumed in SA08 (see Sections \ref{hawki} 
and \ref{protostars} for more detailed information on individual sources).

\begin{figure}
\plotone{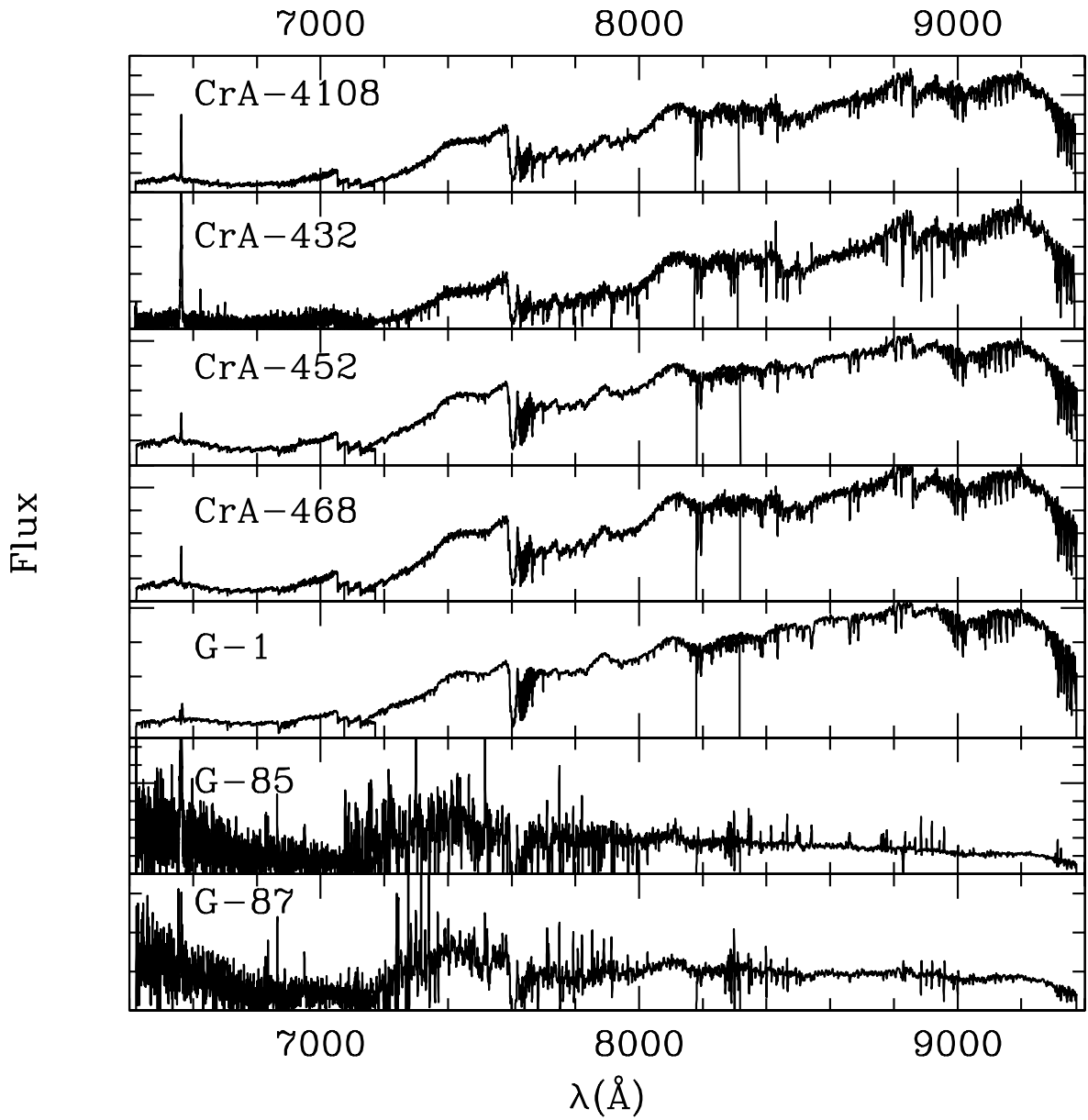}
\caption{FLAMES spectra for the objects with enough S/N to derive a spectral type. \label{spec-fig}}
\end{figure}

\begin{figure}
\plotone{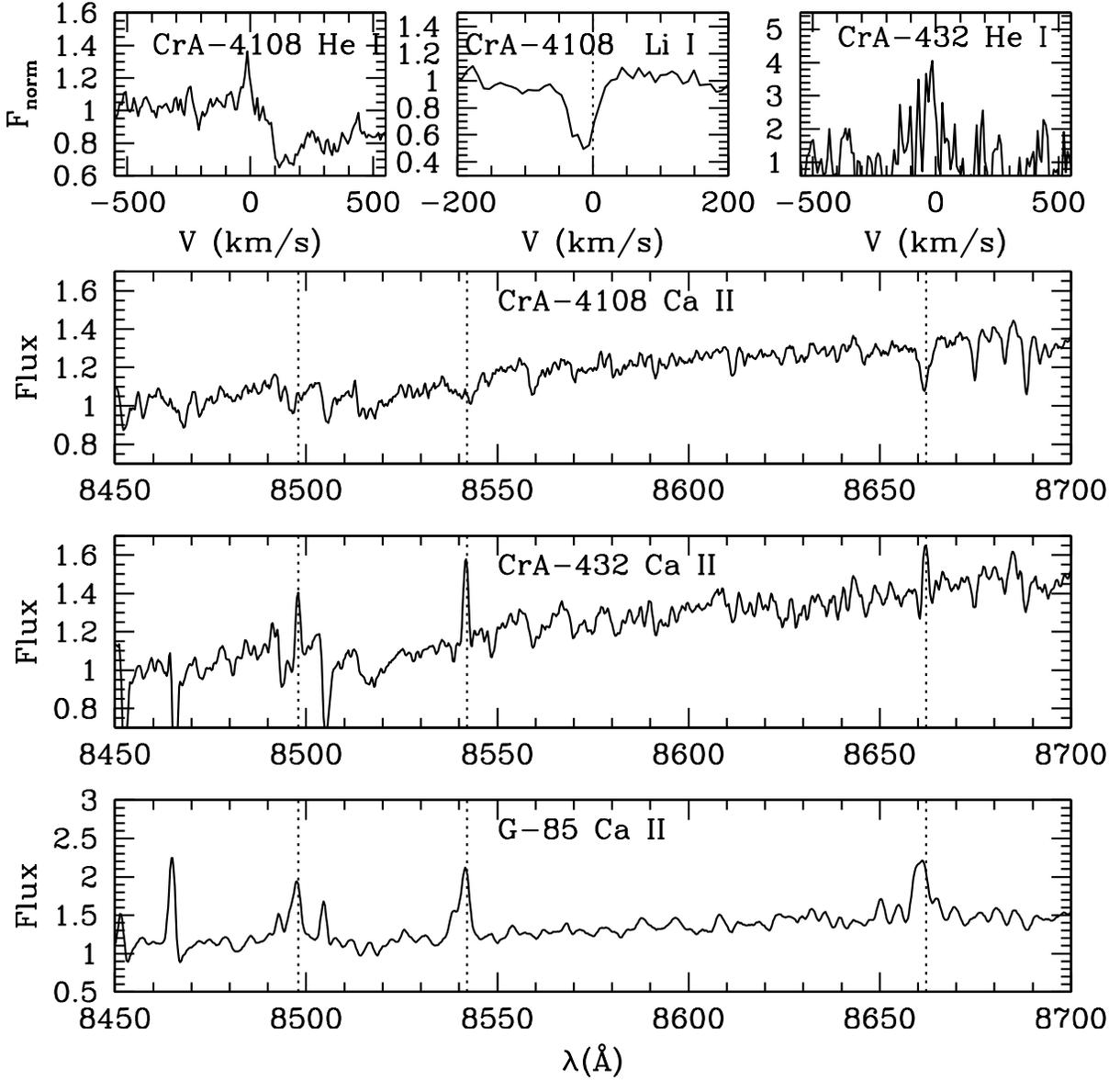}
\caption{Examples of other lines detected in the spectra. The presence of
He I 5875\AA\ and the Ca II triplet in emission are indicators of accretion, Li I is 
a marker of youth. \label{lines-fig}}
\end{figure}

The spectra were reduced using standard tasks within IRAF\footnote{IRAF is distributed 
by the National Optical Astronomy Observatories, which are operated by the Association of
Universities for Research in Astronomy, Inc., under cooperative agreement with the 
National Science Foundation.}. The data were bias corrected
and then extracted with the help of corresponding dome flats, flat fielded,
and calibrated in wavelength using the available ThAr lamp. Individual
spectra were then combined. 
No flux calibration was performed, due to the inherent difficulties of
calibrating absolute fluxes with a multifiber spectrograph and to the
fact that flux calibration is irrelevant for our science case.
Since the sky emission is largely variable throughout the field, 
the sky subtraction was performed in several steps. First, we combined
all the sky spectra for each filter and subtracted from all the
objects. We then examined each object to determine whether this
general sky subtraction properly removed nebular and sky lines.
For the objects located in
the brightest parts of the nebula (CrA-432, G-80, G-85, G-87, G-88,
G-90, G-32, and G-48), for which the standard sky subtraction proved 
insufficient, we then selected the brightest sky
spectra in the region (29 in total) and combined them in a new ``bright'' sky
template, which provided an adequate removal of sky lines. Figure \ref{spec-fig}
shows the high- and moderate-S/N spectra, and 
Figures \ref{halpha-fig} and \ref{lines-fig} provide more details on H$\alpha$
emission and other accretion-related lines.

\subsection{HAWK-I \label{hawki}}

Using the IR camera HAWK-I on the VLT, we observed 11 7.5'$\times$7.5'
fields in the Coronet cluster, as part of our ESO program 083.C-0079.
The location of the fields was determined by the presence of bright objects,
given the minimum magnitude limits of HAWK-I 
(Ks=8.6 mag, H=9.6 mag, J=9.3 mag). Due to these strong magnitude limitations, 
the central part of the cluster and most of the areas with strong star 
formation could not be observed with HAWK-I.
Observations took place on 6 different nights, 2009 April 28 and 29,
2009 June 28 and 30, 2009 July 12 and 16, under clear weather and
variable seeing conditions (0.5-1.0").

The data was reduced using IRAF tasks within the \textit{ccdred} package
for bias correction and flat fielding. For each field and filter, we took
6 dithered images with offsets of 30" and NDIT=20, with total exposure times
of 40s (J), 34s (H), and  58s (K). We used the IRAF \textit{daofind} package
to identify detections, and \textit{apphot} to perform aperture photometry,
with 10 pix aperture and 20-25 pix sky annuli. The photometry was calibrated
field-by-field and chip-by-chip as relative photometry by comparing to 2MASS 
objects present in the fields, using typically 20-120 stars for calibration.
Typical calibration errors range from 1 to 5\%, depending on the weather 
conditions. Therefore, signal to noise and background remain the main sources 
of uncertainty, especially for the fainter objects. The data are complete
in a large dynamical range (J=11-18.5 mag, H=11-18.5 mag, K=10.5-18 mag). 
 
Table \ref{hawki-table} lists the magnitudes of the previously 
known sources detected with HAWK-I. 
Two of the X-ray sources in Garmire \& Garmire (2003) that had not been detected
with Spitzer nor FLAMES (G-116 and G-119) are now detected with HAWK-I.
Their colors suggest that they are most likely extragalactic.
Table \ref{hawki-table} also includes a comment on the variability
of the sources that are also detected in 2MASS. CrA-4107 and CrA-452
show some degree of variability (0.1-0.3 mag, depending on color) which is
probably due to rotation and the presence of spots. 
The J magnitude of G-32 appears more than 1 magnitude fainter in our
observations than in the 2MASS catalog, but since the 2MASS magnitude is
close to the detection limit of the survey, the HAWK-I data are more
accurate.

\subsection{APEX/LABOCA 870$\mu$m mapping \label{apex}}

The 870$\mu$m data were obtained with APEX using the
LABOCA (Large Apex BOlometer CAmera) bolometer array 
as part of the ESO program 081.C-0204(A).
The data consist of 108 scans in raster map spiral mode,
covering an approximate area 30' in diameter. This results
in an integration time of 9.2 hours on target. 
The observations were carried out between 2008 June 21 and 2008 
August 24. The opacity was determined with skydips performed at
the beginning and the end of the night, varying between $\tau$=0.15-0.44,
with a median value 0.2. The flux calibration was estimated
observing several sources per night, including Neptune, Uranus,
and the secondary calibrators J1957-3845 and J1924-292. 

\begin{figure}
\rotate
\includegraphics[scale=1.0,angle=0]{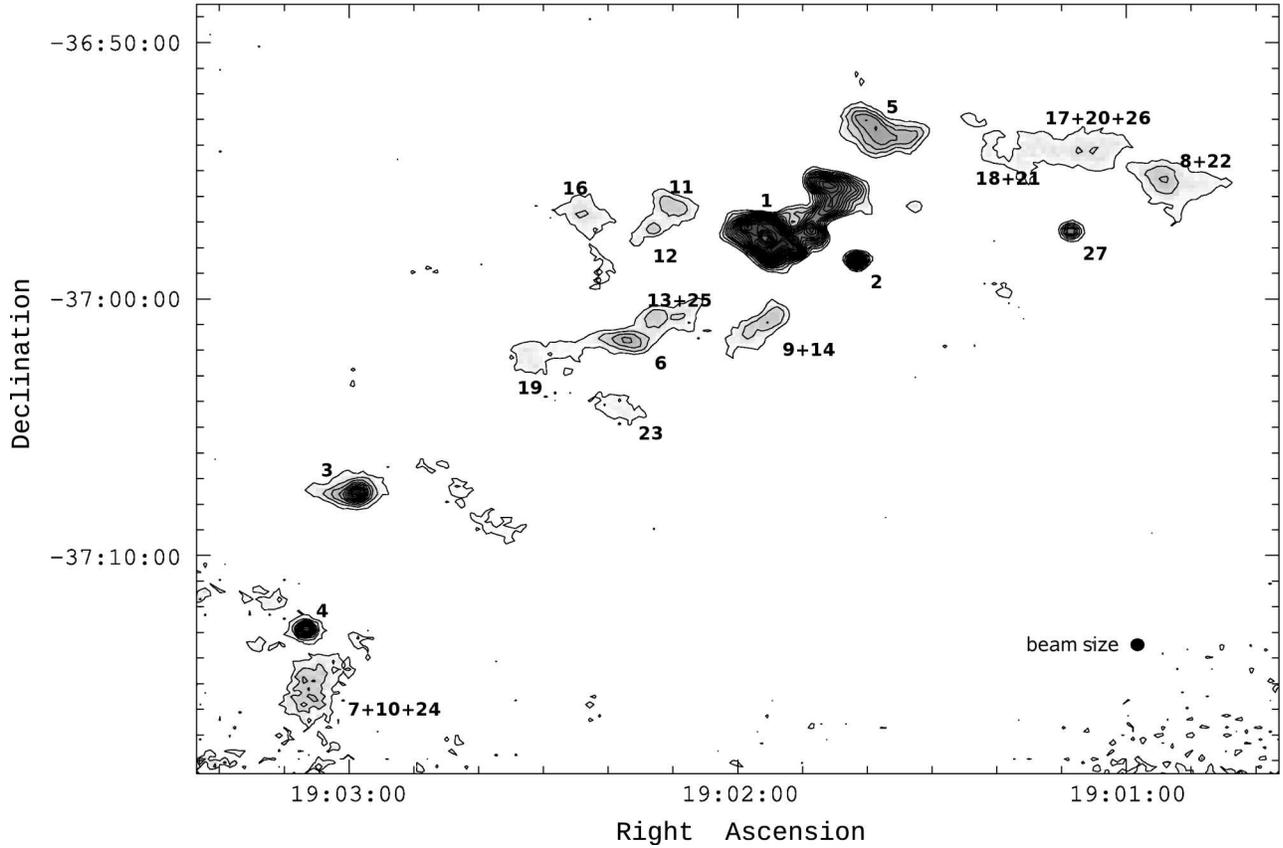}
\caption{Submillimeter  870$\mu$m map obtained with APEX/LABOCA. The greyscale
represent a logarithm scale between 0.030-1 Jy (approximately 
3$\times$rms). The contours are in lineal scale, from 0.030 to 1 Jy
 in steps of 0.050 Jy. The beam is displayed in the lower right corner.
The numbers correspond to the clumps identified by Clumpfind 2D (see also 
Table \ref{laboca-tab}). Right ascension and declination are in J2000.\label{laboca-fig}}
\end{figure}

The map was created and calibrated following the pipeline reduction
of LABOCA data within the BoA package, especially developed for 
APEX bolometer data \footnote{See details in 
http://www.astro.uni-bonn.de/boawiki/Boa}. BoA allows to reduce skydips to 
derive the opacity, reduce the calibrators, and make the maps. The 108 
individual maps were reduced separately, and then coadded to obtain the 
final map. In each map, the data affected by high acceleration were 
flagged, as well as the dead and very noisy channels (with
rms larger than 5 times the median). Correlated
noise was removed by comparing groups of channels. Finally, a linear
baseline was fitted and subtracted from the data. The final maps
were obtained using the $weakmap$ option, which is optimized for
weak sources and also provided the best result for the extended
emission in the region. The final rms in the map center is $\sim$0.010 Jy, 
being higher near the brightest sources and the less-integrated edges
of the map. The resulting map is displayed in Figure \ref{laboca-fig}.

\begin{figure}
\rotate
\plottwo{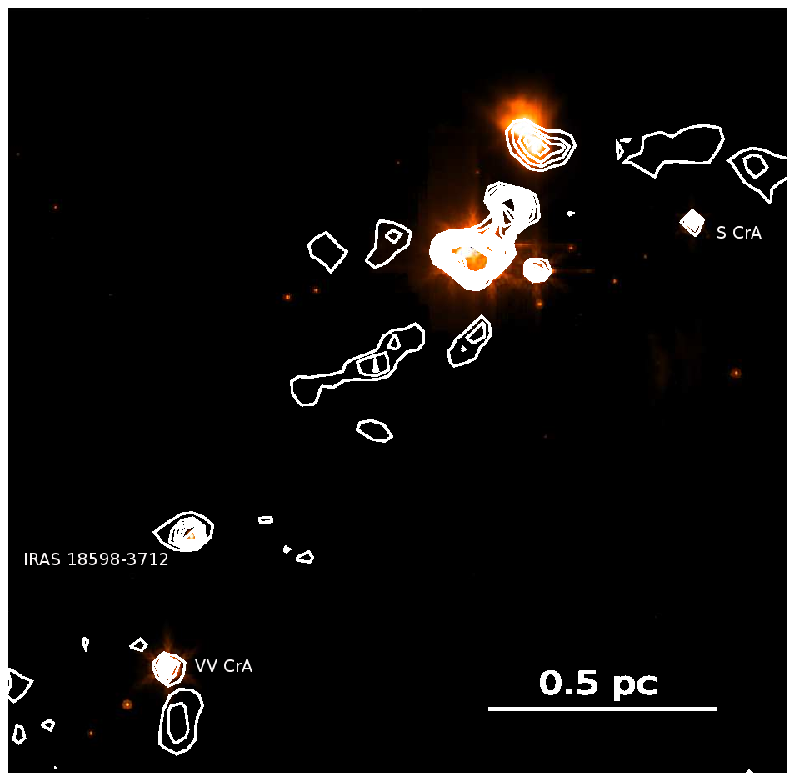}{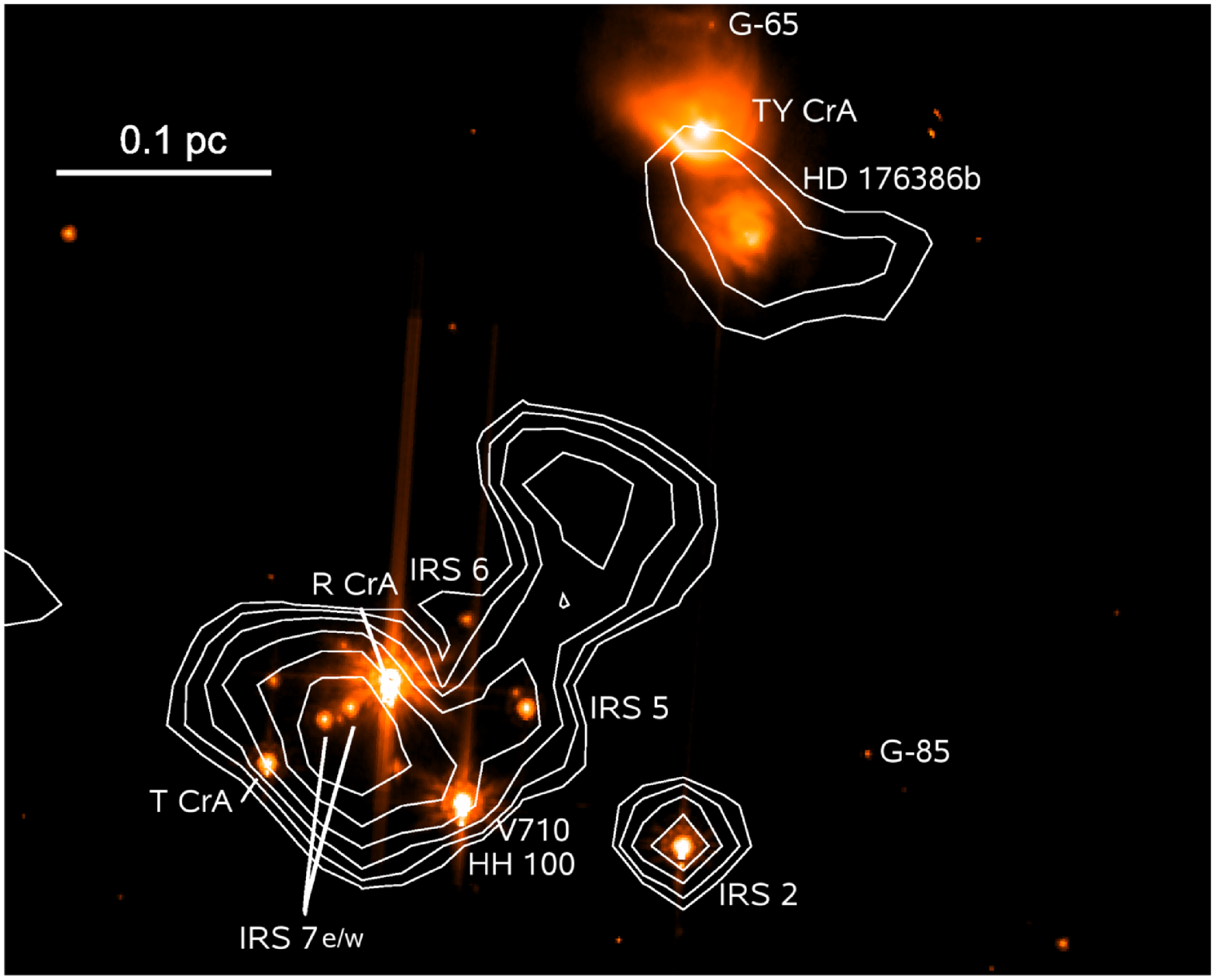}
\caption{Left: LABOCA contours for the detected clumps, plotted over 
the MIPS 24$\mu$m image of the coronet cluster. Right: A zoom in the densest part
of the nebula. The scale bars assume a distance of 170 pc. \label{contours-fig}}
\end{figure}

The second step was to identify and extract the detected structures
in the map and to measure the fluxes. The cluster contains a large,
dense structure with several peaks as well as some elongated structures
and isolated clumps (see Figure \ref{laboca-fig}). The identification and extraction
of clumps in the elongated and isolated structures was done using 
the 2d Clumpfind algorithm, developed by Williams et al. (1994).
The 2d Clumpfind works by contouring the data according to user-defined
levels in order to determine the presence of structures.
To efficiently detect the clumps in the different parts of the map,
which have in general different sensitivity, we defined levels according
to the local rms. The most noisy parts of the maps were removed
before running Clumpfind. The clump identification was repeated 5 times
with various sets of levels, and only clumps detected in 3 or more
trials are considered to be real. Several sources appear as a
single clump or as a collection of adjacent clumps depending on the
search parameters. Given the uncertainty on the source boundaries and
fluxes, they are listed as a ``multiple source'' with a unique flux
in Table \ref{laboca-tab}, with names that indicate that they are composed
of more than one peaks or individual sources (7+10+24, 8+22, 9+14, 13+25,
17+20+26, 18+21).
The flux for each clump (or clump collection) was calculated with
$clstats2d$, also within the 2d Clumpfind package. The errors in the final
flux include the local rms as well as the standard deviation from the flux measured
by the slightly different levels imposed for the source detection. The final clumps are
listed in Table \ref{laboca-tab}. In general, small, compact structures are detected
in a very robust way, while the limits and subdivisions of more extended structures are
less well constrained. Several of the submillimeter sources are
associated to known IR sources and/or nebular structures (Figure \ref{contours-fig}).

\section{Analysis \label{analysis}}

\subsection{Extinction maps and submillimeter observations\label{maps}}

We used the near-infrared colors of the sources detected in the HAWK-I  
frames to calculate the dust extinction through the CrA cloud. In  
particular, we adopted the color-excess mapping method \textsf{nicer}  
\citep[][]{lom01}. In the following, we briefly describe the practical  
implementation of the method. For further details of the method we  
refer to \citet[][]{lom01} and \citet[][]{lad94}.

The observed colors of stars shining through a dust cloud are related  
to the dust extinction via the equation:
\begin{equation}
E_\mathrm{i-j} = (m_i - m_j) - \langle m_i -m_j\rangle_0 =  
A_j(\frac{\tau_i}{\tau_j}-1),
\label{eq:colorexcess}
\end{equation}
where $E_\mathrm{i-j}$ is the color-excess, $A_j$ is extinction, $ 
\tau_j$ is optical depth, and $\langle m_i -m_j\rangle_0$ refers to  
the intrinsic color of stars, measured as the mean color of stars in a  
field that is free from extinction. We calculated the mean colors  
using stars in a HAWK-I frame close to ($RA$, $Dec$ J2000) = (19:02:50,  
-36:46:25), estimated to contain very little extinction based on our  
earlier, lower-resolution extinction map of the CrA region  
\citep{kai09}. The mean colors in this field were $\langle m_H -m_K 
\rangle_0 = 0.19 \pm 0.13$ mag and $\langle m_\mathrm{J} -m_\mathrm{H} 
\rangle_0 = 0.47 \pm 0.15$ mag. For the ratios of optical depths we  
adopted $\tau_H / \tau_K = 1.67$, $\tau_J / \tau_H = 1.48$, and $ 
\tau_V / \tau_K = 8.77$ \citep{car89}. Before calculating the  
extinction, we removed from the HAWK-I detections the previously known  
members of the CrA cloud, based on the identifications given by SA08, 
Forbrich \& Preibisch (2007), and L\'{o}pez-Mart\'{\i} et al. (2005, 2010).

Observations in three broadband filters ($JHK_\mathrm{S}$) result in  
measurements of two color excesses, namely $E_\mathrm{H-K}$ and $E_ 
\mathrm{J-H}$, towards each source. These two measurements were  
combined to yield extinction values towards the sources following the  
formalism presented in \citet{lom01}. The derived extinction values  
were then used to produce a regularly sampled map by computing the  
mean extinction in regular intervals (map pixels) of $24\arcsec$,  
using a Gaussian with $FWHM=48\arcsec$ as a spatial weighting  
function. In this resolution, the maximum extinction in the resulting  
map reaches $A_\mathrm{V}\approx 50$ mag, with $A_\mathrm{V}\approx 20$ mag 
corresponding to 1-2 stars per pixel. The error in the map depends  
on extinction, with 3-$\sigma$ being about 0.8 mag at $A_\mathrm{V} =  
0$ mag and 7 mag at $A_\mathrm{V} = 50$ mag.

\begin{figure}
\plotone{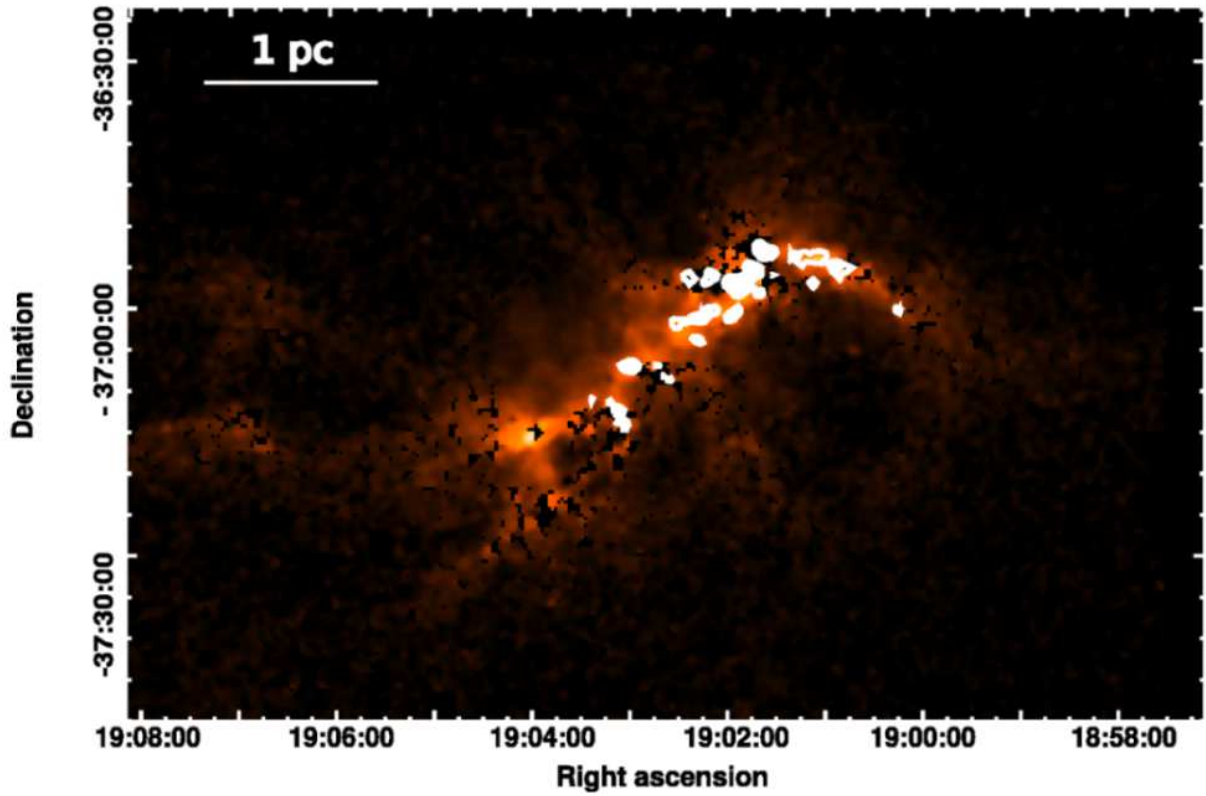}
\caption{Extinction map and submillimeter structure (see text). Right ascension and
declination are in J2000. The scale bar for 1 pc assumes
a distance of 170 pc. The color gradient represent the extinction in a log scale
from A$_V$=0.8 to 50 mag. \label{extinction-fig}}
\end{figure}

Figure \ref{extinction-fig} shows the resulting extinction map, 
combined with the 2MASS
extinction map in the regions not included in our HAWK-I imaging. There
is an excellent agreement between the 870$\mu$m clumps and the regions
of high extinction. Nevertheless, the extinction map reveals large parts
of the molecular cloud that are not detected in the LABOCA survey due to
their lower column density. The most massive stars in the region are associated
to the zones with the highest extinction and strongest submillimeter emission,
as expected. The distribution of low mass members with respect to the cloud
is less clear, also due to the fact that the non-biased survey for low-mass
members was done only in the central part of the cluster.

\subsection{Spectral types, extinction, and accretion\label{spectype}}

The FLAMES spectra allow to determine spectral types for the objects 
with good S/N. We followed a similar scheme as used in SA08, using
the spectral indices and bands from Mart\'{\i}n et al. (1996), called PC1, PC2, PC3 
and PC4, and the R1, R2, R3, TiO~8465 indices in Riddick et al. (2007).
The additional indices VO~2 and VO~7445 from Riddick et al. (2007) were also used to
refine the spectral types for objects M5 or later. Table \ref{index-table} offers
a list of the relevant indices and their valid spectral type ranges. A combination of
spectral types and the available JHK photometry allows to estimate
the extinction of the objects (see below). For the objects with intermediate
to high extinction (A$_V >$3 mag), we gave priority to the
results derived from the indices PC1, PC2, R1, R2, R3, TiO~8475, VO~2, and VO~7445,
which are derived from flux rations in adjacent parts of the spectrum, being
thus less sensitive to extinction. For objects like CrA-432, that have
nearly no signal in the bluer part of the spectrum (L682.2), we used only the indices
involving the longer-wavelength parts. For every object, the spectral types
derived from each index were compared, and those that largely deviated from
the typical value were removed and not used (deviating indices are usually affected by
noise, atmospheric features, and/or poor S/N in one of the bands). In 
general, between 2-6 indices were used for the classification of each star, depending
on the spectral type, extinction, and S/N in the three spectral bands. 
The errors in the spectral types are calculated as the standard deviation of
the values derived from each valid index. The 
classification scheme failed for the object G-1, which appears earlier than the
rest. It was thus classified by comparison to solar-type T Tauri star templates 
obtained from a combination of
Hectospec/MMT observations in the Cep OB2 region (Sicilia-Aguilar et al. 2005).

\begin{figure}
\plotone{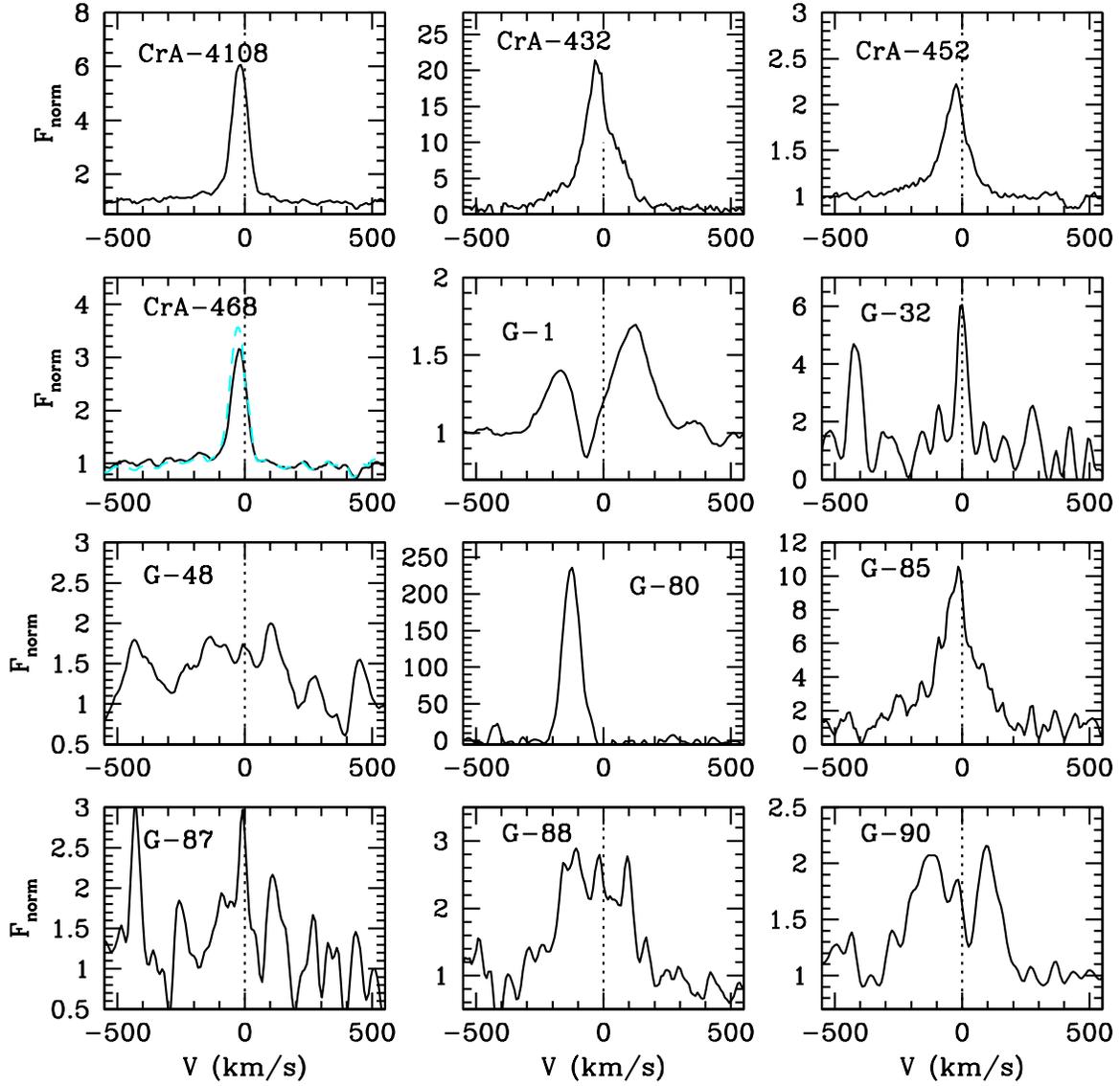}
\caption{Normalized spectra in the H$\alpha$ region for the detected objects. 
For CrA-468 (G-49), we display as a dashed cyan line our previous FLAMES spectrum (SA08).
 \label{halpha-fig}}
\end{figure}

\begin{figure}
\plotone{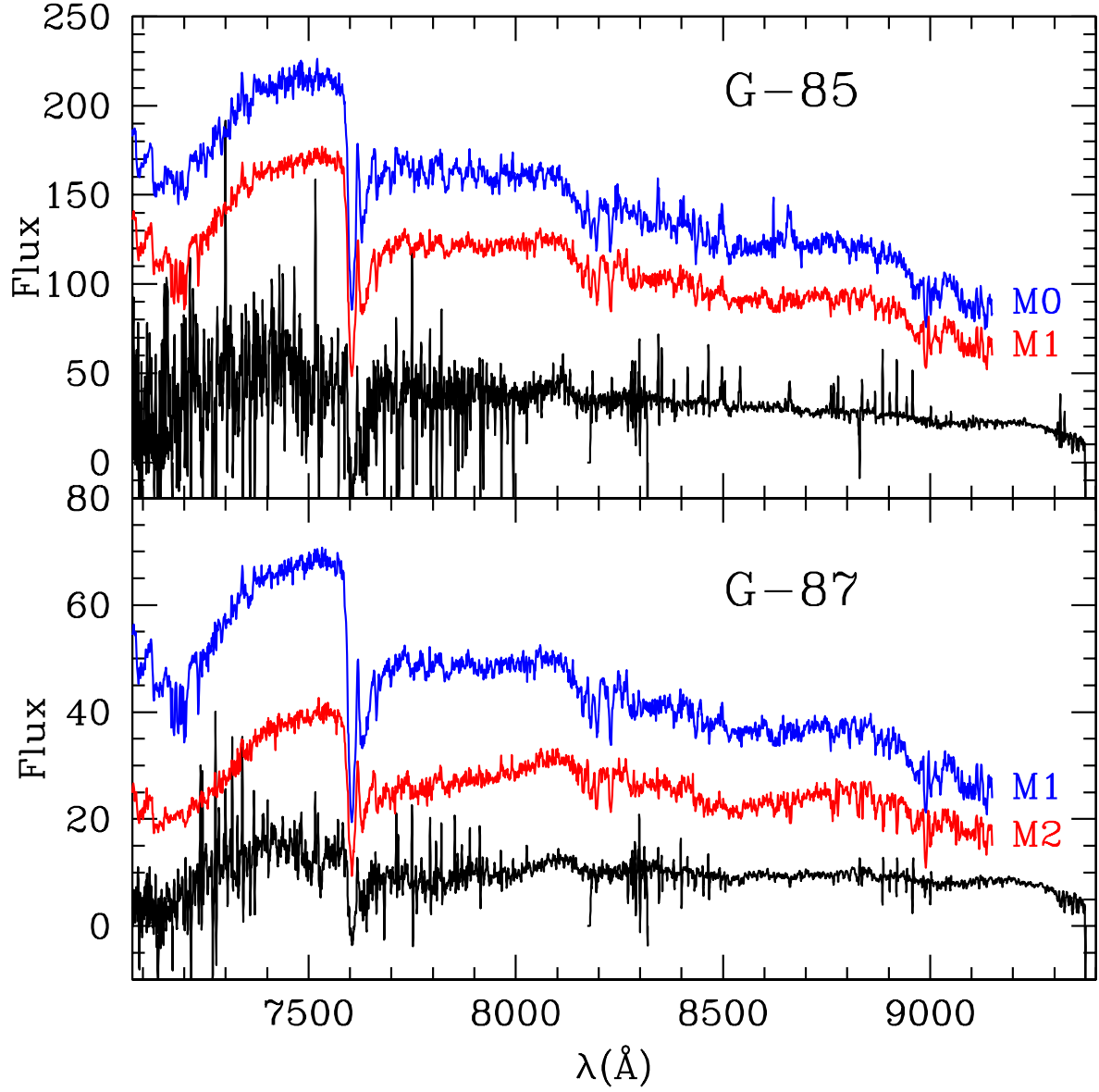}
\caption{Dereddened spectra of the low S/N objects G-85 and G-87, compared to our spectral type
templates. \label{G8587-fig}}
\end{figure}

Two of the objects,
G-85 and G-87, are highly extincted (A$_V >$14 mag). After a first tentative classification,
we used the preliminary A$_V$ value to de-redden the spectra (using IRAF task $deredden$
within the $onedspect$ package) before the final classification.
Examination of two objects revealed that their
spectral types are earlier than M2, a range for which the mentioned indices
are not valid. We thus derived their spectral types using our solar-type
T Tauri templates as we did for G-1 (see Figure \ref{G8587-fig}).
Note that G-85 and G-87 had been previously observed in SA08, but the
new improved S/N (due to better weather and seeing conditions) now allows to refine
their spectral types, confirming them to be earlier than previously thought
by about 2-3 subtypes (having G-85 and G-87 spectral types M0.5$\pm$1.5 
and M1.5$\pm$1.5, respectively). The star CrA-468, also known as G-49, had also
previously been observed with FLAMES (SA08). The new spectrum is
fully consistent with the previous one (see Figure \ref{halpha-fig}),
resulting in a similar H$\alpha$ classification and spectral type 
(M4.0$\pm$1.0 versus M3.5$\pm$0.5).
The final spectral types are listed in Table \ref{lines-tab}.

The individual extinction values were derived from the JHK 2MASS photometry, comparing to 
standard colors for stars and BD (Bessell et al. 1998; Kirkpatrick et al. 1995).
The extinction was derived from E(J-H) and E(H-K), using the relations in
Cardelli et al. (1989), A$_V$=10.87 E(J-H) and A$_V$=5.95 E(J-K) mag,
valid for a standard galactic extinction law. For most sources, the agreement
between A$_V$ obtained from the two different colors suggests that most of 
the extinction is due to foreground and cloud material, with no evidence of anomalous extinction
that could be caused by processed grains in the circumstellar environment,
despite the presence of strong disk emission in
some of the objects. Only in the cases of CrA-4108 and CrA-468 we found
large differences between the extinction values derived from different colors,
which could indicate the presence of anomalous extinction/scattering. In the case of
CrA-468, the Spitzer data reveals it is located near substantial
nebular material. CrA-4108 is out of our Spitzer field.

The presence or lack of accretion was established by measuring the
equivalent width and examining the profile and line width at 10\% of the peak
for the H$\alpha$ line (see Figure \ref{halpha-fig}).
The lines were measured using IRAF tasks within the $splot$ tool.  
We adopted the classification of White \& Basri (2003) for the limiting EW
of accreting and non-accreting stars, depending on their spectral type.
Given that the line width is a more powerful tool than 
the EW to detect weak accretors (Sicilia-Aguilar et al. 2006), we consider
an object to be accreting if its 10\% line width exceeds 200 km/s (Natta et al. 2005). 
In some cases, like G-1, the broad profile is evident, although the 
EW is much smaller than the canonical limit for accretion in an M0.5
object, so we list it as an accretor. The presence of accretion was also checked in other
lines, like the Ca II IR triplet and the 6678\AA\ He I line.
The lack of Ca II emission in G-1 suggests an
accretion rate below 10$^{-9}$ M$_\odot$/yr (Muzerolle et al. 1998).
All the spectral lines measured are listed in Table \ref{lines-tab}.

The new accretion determination, together with our previous
measurements in SA08, confirms that most of the low-mass Coronet members
have very low accretion rates, with CrA-432 and G-85 being
the only exceptions. As we found in SA08, transitional and highly
evolved disks (like G-87) do not show evidence of ongoing accretion (within the 
10$^{-10}$-10$^{-11}$ M$_\odot$/yr limit imposed by the FLAMES spectral resolution).
This differs from the observations of solar-type stars,
for which usually half or more of the transition disks are found to be accreting
(Sicilia-Aguilar et al. 2006, 2010; Muzerolle et al. 2010; Fang et al. 2009).
All objects without IR excesses are consistent with no accretion.
For the objects out of the Spitzer fields (CrA-4108 and CrA-452),
the spectroscopy reveals no evidence of accretion. CrA-4108 shows marginal
emission that could belong to He I at 5875\AA, but the lack of Ca II
lines in emission suggests that the object is not accreting (see 
Figure \ref{lines-fig}). The Herbig Haro (HH)
object G-80 was reobserved in this program, confirming again its shock
nature via [N II], [S II], and asymmetric H$\alpha$ profile. Finally,
three sources that were not detected with Spitzer (G-48, G-88, and G-90)
are now marginally detected, with G-88 and G-90 displaying broad
H$\alpha$ emission that suggests they are very embedded Class I sources.
Their condition as embedded sources is also consistent with their location 
within one of the strongest submillimeter structures.

\subsection{Massive disks, protostars, and prestellar condensations \label{protostars}}

\begin{figure}
\plotone{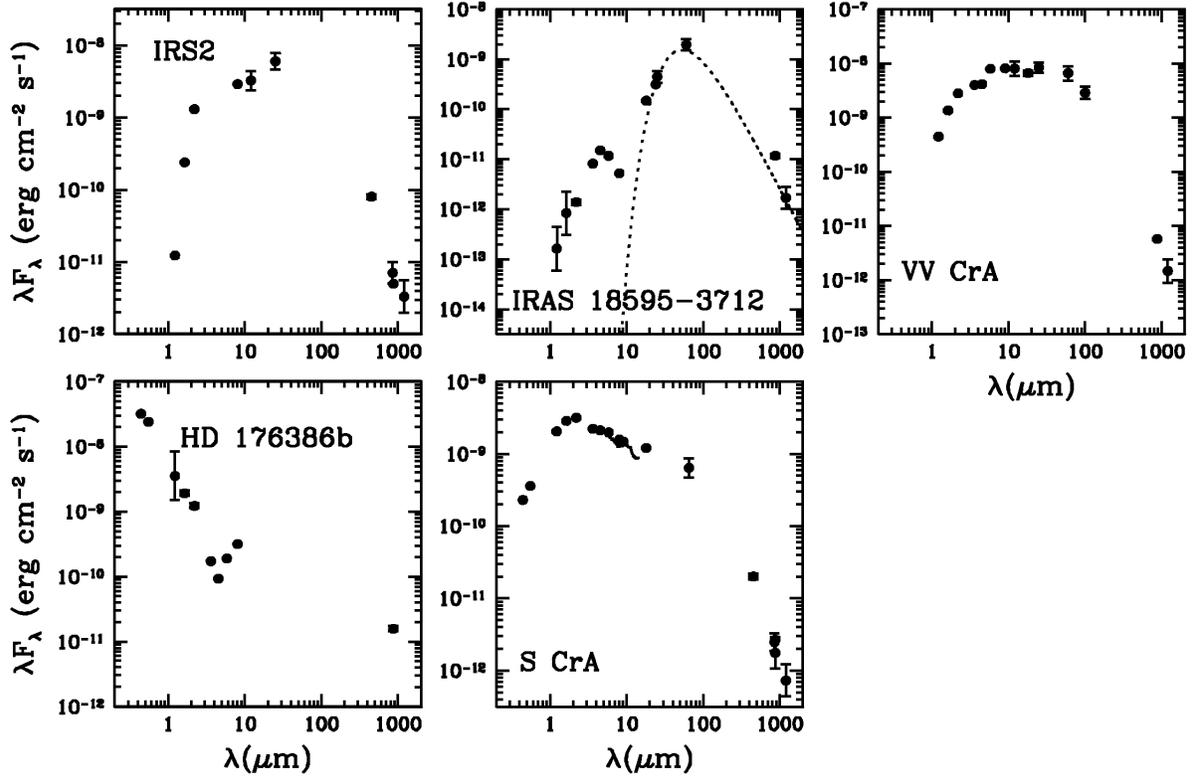}
\caption{SEDs of the unresolved sources detected with
APEX/LABOCA. The SEDs include the data listed in Table \ref{seds-table}. The data for S~CrA is corrected for extinction, taking
A$_V$=0.4 mag (Currie \& Sicilia-Aguilar 2011). A scaled black body emission for T=70 K is plotted as
a comparison for the long wavelength emission of IRAS 18595-3712. \label{seds-fig}}
\end{figure}

As listed in Table \ref{laboca-tab}, several of the detected
870$\mu$m sources are related to known Class II and embedded
sources in the region. Assuming a distance of 170 pc, the
18.6" beam results in a $\sim$3000 AU size. Condensations \#2, 3, 4, 5, and 27 are
associated to the previously known sources IRS~2, IRAS~18595-3712, VV~CrA, 
HD~176386b, and S~CrA, respectively. In addition, \#3 and 4 also correspond
to the millimeter emitting objects 23 and 24 from Chini et al. (2003).
We combined the pre-existing data from Spitzer (see Currie and Sicilia-Aguilar
2011 for the data on IRS~2, S~CrA, and HD~176386b and the reduction procedures applied
to the rest of the sources),
AKARI (Kataza et al. 2010), and millimeter/submillimeter data
(Gezari et al. 1999; Chini et al. 2003; Nutter et al. 2005), to trace the SEDs
of these objects. Table \ref{seds-table} summarizes the available data for
the SEDs shown in Figure \ref{seds-fig}.

The 870 $\mu$m flux from IRS~2 is in good agreement with previous
submillimeter measurements of the source, and consistent with
an embedded Class I source. For S~CrA, our data is also in very good
agreement with previous submillimeter observations, and its
SED is consistent with a massive protoplanetary disk. Both IRS~2 and S~CrA appear as
marginally spatially extended, which suggests the presence of 
envelopes or surrounding cloud material. In both cases, the estimated size
of the envelopes would be of the order of $\sim$5000 AU (assuming a distance
of 170 pc; Knude \& Hog 1997). 
The emission associated to HD~176366b is extended, and
appears related to the warm cloud material detected in the Spitzer images.

To the south of the CrA region, two condensations, associated to the
sources VV~CrA and IRAS~18595-3712, are detected. The emission 
associated to the binary VV~CrA is very compact 
and probably corresponds to the protoplanetary disk of the main source plus 
that of the disk and envelope of the embedded companion (Ratzka et al. 2008). 
The submillimeter emission around 
IRAS~18598-3712 appears slightly extended to the East, suggesting the
presence of a larger envelope. The emission at longer wavelengths for this source
is indeed consistent with a black body of T=70 K (see Figure \ref{seds-fig}).
The extinction map reveals substantial
cloud material in this direction, which could also produce the extended
emission near the source. The SED of this object suggests the presence
of a large amount of material, which could include the source envelope
plus contamination from the surrounding cloud.

Regarding the rest of clumps detected in the LABOCA maps, they
correspond to more complex regions that include several embedded
sources and to denser nebular material in the
region, which is confirmed by the extinction maps. 
This is the case of clump \#1, which includes the sources T CrA, R CrA, V710, 
and IRS~7e/w,among others (see Figure \ref{contours-fig} right).
The peak of the emission appears to the South-West of IRS~7,
which is consistent with the observations of Chini et al.(2003) and Nutter
et al. (2005). The embedded X-ray source G-17 is associated to clump \#12.
Although the size of the clump is too large for it to be considered
as the same and single object, the presence of this X-ray source near the 
emission peak of clump \#12 indicates that G-17 is most
likely a very embedded low-mass protostar, as it had been
suggested by SA08. The star CrA-465 appears superimposed to the clump \#9, but 
the submillimeter emission is most likely due to background cloud material not
related to the source, given both the size of the structure
and the low extinction of CrA-465 (A$_V$=0.08$\pm$0.04; SA08).  

The strongest emission in the LABOCA map is thus associated to the
largest concentration of intermediate-mass members, as it would 
be expected. The most massive stars appear thus along the
two arms (running in the HD 176386 to VV CrA direction, and in the HD 176386 to
S CrA directions, respectively) of the molecular cloud, clearly depicted in the
extinction maps (Figure \ref{extinction-fig}). It is not possible to establish 
a further correlation between the rest of structures with
submillimeter emission and the distribution of cluster members
in the region, since so far only the most central part of the cluster
has been explored for membership in a systematic way. 

\section{The age of the Coronet cluster and the star forming region \label{age}}

The age of the Coronet cluster has been subject to intense debate.
While most observations suggested an age of 1-3 Myr (Meyer \& Wilking 2009),
the presence of a Class 0 source and several embedded objects 
was rather consistent with a younger cluster, $\sim$0.5-1 Myr (Henning et al. 1994;
Chen et al. 1997; Chini et al. 2003;  Nisini et al. 2005; Nutter et al. 2005;
Groppi et al. 2004, 2007). Nevertheless, both the disk frequency
($\sim$50-65\%) and the disk characteristics among the low-mass, M-type population
appear closer to the values observed in older regions than to a 1 Myr-old
cluster (SA08; L\'{o}pez-Mart\'{\i} et al. 2010; Currie \& Sicilia-Aguilar 2011).
On the other hand, the presence of substantial amounts of nebular material 
would be inconsistent with an age beyond 3 Myr (Hartmann et al. 2001; 
Ballesteros-Paredes \& Hartmann 2007). Finally, spectroscopic data
also reveal a low accretion fraction, with all the transition and 
settled/dust depleted disks
showing no evidence for accretion, which has been typically 
related to few-Myr old regions (Muzerolle et al. 2010).

Few-Myr age differences between cluster members are common in many
regions, and many clusters include sequential star formation,
which results in the presence of more than one star-forming event
and thus two or more populations with different ages (e.g. Sicilia-Aguilar
et al. 2005, 2006). Nevertheless, the Coronet cluster is special
because of it compactness: the members studied here are located within
the densest region of the CrA nebula, which is less than 1.3 pc across
(adopting a distance of 170 pc; Knude \& Hog 1997). A strong correlation between 
the size of the region and the age dispersion of the cluster
members has been observed in our galaxy and in the Magellanic Clouds
(Elmegreen 2000): the age dispersion of cluster members is of
the order of the crossing time in the region, which for the
Coronet cluster is smaller than 2 Myr. In fact, regions similar to
the Coronet cluster in size and stellar content (e.g. the $\eta$ Cha
cluster) are usually found to be strongly coeval (Lawson et al. 2009).

\begin{figure}
\plotone{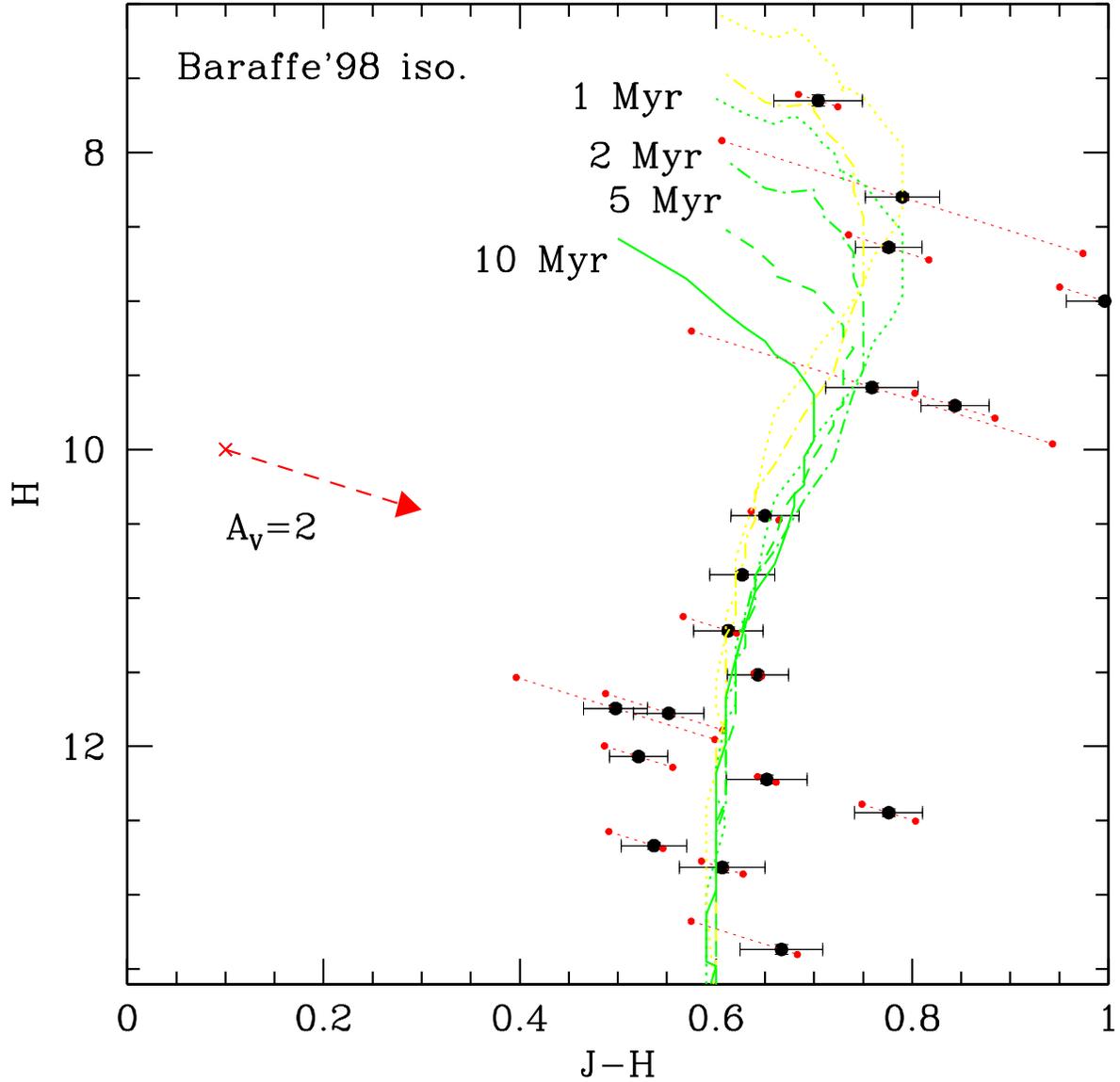}
\caption{Color-magnitude diagram for the
T Tauri stars in the Coronet cluster, including the new objects classified in
the present work and those from our previous spectroscopy run presented in
SA08. The error bars of each point include the photometry error only;
in addition we add a reddening vector per data point displaying the uncertainty
in the extinction/spectral type. The Baraffe et al. (1998) isochrones are displayed for two 
distances: 130 pc (yellow; Marraco \& Rydgren 1981) 
and 170 pc (green; Knude \& Hog 1998). \label{colormagiso-fig}}
\end{figure}

We have thus estimated the age of the cluster by placing the 
known members in a color-magnitude diagram. 
A color-magnitude diagram is preferred to a HR diagram since it involves
less transformations of the original data that result in increased 
uncertainties. The effect of the data uncertainties is also easier to
control and determine in a color-magnitude diagram, since it does not involve propagation
of non-independent errors.
Our knowledge of spectral
types allows to correct individually for the extinction, which is
particularly important in regions with non-uniform cloud material.
Figure \ref{colormagiso-fig} displays the near-IR color-magnitude diagram
H vs. J-H of the low-mass T Tauri members of the Coronet cluster, 
using their 2MASS and HAWK-I data, together with the
isochrones from Baraffe et al. (1998) for the distances
of 130 (only the 1 and 2 Myr isochrones) and 170 pc (1, 2, 5, and 10 Myr
isochrones). These two distances correspond to the estimates given by 
Marraco \& Rydgren (1981, 130 pc) and the more recent one by Knude \& Hog (1998,
170 pc). Given that age and distance are degenerated, we consider both extreme
values in order to ascertain the potential errors in the age determination.
Table \ref{jhk-table} summarizes the data of the sources used in the
HR diagram. The extinction of the objects has been corrected using
their individual values in Figure \ref{colormagiso-fig}. In addition to the photometric errors, we also
display the uncertainties in the extinction, which are related to 
the uncertainties in the spectral types of the objects.
Due to the fact that there is little spread for the lower-mass objects
in the J-H direction at these ages, it is hard to determine an
age for the cluster from these objects. Nevertheless, the location
of the isochrone ``kink'' for the early M-type stars is more consistent
with a distance of 170 pc. In any case, for either a distance of 130 or 170 pc,
and considering the limitations imposed by the small number of objects
with early M-type in the cluster,
the distribution in the color-magnitude diagram of the early M-type objects 
suggests that the population has an age in the range 0.5-2 Myr. For
the most plausible distance of 170 pc, the age would be in the 0.5-1 Myr range. 
This is in agreement with the conclusions of Chen et al. (1997), based on
the bolometric temperature and luminosity of young stellar objects
(YSO), which stated that the CrA region is younger than 2 Myr and
most likely, younger than Taurus, even assuming a distance of 130 pc.

This age estimate is also consistent with the presence of Class 0 and
very embedded Class I sources in the region. Nevertheless, the 
relatively low disk fraction found among T Tauri stars (2/3 or lower; 
SA08, L\'{o}pez-Mart\'{\i} et al. 2010) is atypical for such a young
region. The ``evolved'' features of the disks (lack of silicate emission
as expected from disks with large grains, presence of disks with inner holes and
small dust-depleted disks, SEDs consistent with strong dust
settling; SA08, Currie \& Sicilia-Aguilar 2011)
are hard to understand for a 1 Myr-old cluster. 
Some of the objects (CrA-205, CrA-4108, CrA-4109) are located at larger
distances from the center of the cloud, and may have thus a larger
age difference with the main cluster. Nevertheless, the lack of 
spectral types and extinction (for CrA-205 and CrA-4109) and the 
position of the source itself in a place where the isochrones are
degenerated (for CrA-4108) do not let us test this hypothesis.
Although the Coronet is
younger than (or has a similar age to) Taurus, its median disk SED
for M-type stars is well below the Taurus median SED for the same type
of objects (SA08). This may trace back to
the formation history of the region and/or the evolution of disks
around these low-mass objects in relatively quiescent and sparse
regions. It could also be explained if the cluster were unusually compact
(and actually not sparse) and interactions would affect the disk structure
and evolution. Further observations in a more extended area would be required
to check for the presence of other populations.

\section{Conclusions \label{conclu}}

We present new multiwavelength data for the Coronet cluster,
including optical spectroscopy obtained with the FLAMES multifiber spectrograph,
870$\mu$m mapping obtained with the LABOCA bolometer on the APEX telescope, 
and near-IR imaging from HAWK-I. The optical data allows to 
derive spectral types, individual extinctions, and accretion status for the remaining 
X-ray members, completing the work started by SA08. The submillimeter
data reveals the densest structures of the cloud, as well as the 
circumstellar disks and envelopes of some intermediate-mass and T Tauri
members. The near-IR imaging provides photometry for some of the faintest members and
allows to derive extinction maps, which trace the cloud material 
in low-density regions that are not detected in the submillimeter.
Both extinction and submillimeter maps reveal a large cloud where star
formation has only occurred along two arms that contain the densest parts of the
molecular cloud, resulting in a quite compact cluster.

By placing the extinction-corrected photometry of the known low-mass members 
in a near-IR color-magnitude diagram and comparing with the Baraffe et al. (1998)
isochrones, we estimate the age of the cluster to be below 2 Myr, and
probably within the 0.5-1 Myr range. This agrees with previous studies 
(suggesting an age $<$3 Myr; Meyer \& Wilking 2009) as well as with the 
presence of substantial cloud material and highly
embedded Class I and Class 0 objects, and it is also in agreement with the 
compactness of the region where most members are located. 
But a young age is in a strong contrast with the relatively low disk fraction 
($\sim$65-50\%; SA08; L\'{o}pez-Mart\'{\i} et al. 2010) and the
evidence of disk evolution (inner holes, lack of silicate features, strong settling,
and dust depleted disks; SA08; Currie \& Sicilia-Aguilar 2011)
observed among the lower-mass and M-type cluster members. Further observations will
be required to examine the presence of other low-mass members in the region
and determine whether the region contains more than one population.

Acknowledgments: We want to thank J. Rod\'{o}n and M. Schmalzl for
their help with the CLASS and Clumpfind routines,
B. Ziegler for his careful assistance in the ESO service-mode observations,
M. Fang for his help with the figures, and the anonymous referee for his/her 
comments that helped to clarify this paper.
A.S.-A. acknowledges support by the Deutsche Forschungsgemeinschaft (DFG)
grant number SI-1486-1/1.
This paper contains the results of the ESO programs 083.C-0079 and 080.C-3024.
This work makes use of data products from the Two Micron All Sky Survey, which is 
a joint project of the University of Massachusetts and the Infrared Processing and Analysis 
Center/California Institute of Technology, funded by the National Aeronautics and Space 
Administration and the National Science Foundation. This research has made use of 
the SIMBAD database, operated at CDS, Strasbourg, France.

\clearpage

\begin{deluxetable}{lccccccccccccl}
\tabletypesize{\scriptsize}
\rotate
\tablenum{1}
\tablecolumns{13} 
\tablewidth{0pc} 
\tablecaption{Spectral types, extinction, and main lines \label{lines-tab}} 
\tablehead{ \colhead{Name} &  \colhead{ID/Coord.} &\colhead{Spec. Type}  & \colhead{A$_V$ (mag)} & \colhead{H$\alpha$} &  \colhead{Li I} & \colhead{[N II]} & \colhead{[S II]} & \colhead{He I} & \colhead{Ca II} & \colhead{K I} & \colhead{Na I$^R$} & \colhead{Na I$^I$}  & \colhead{Comments} } 
\startdata
CrA-4108 & 19020968-3646345	& M3.5 ($\pm$1.0)   &  1.2$\pm$1.1   & -7.6 & 0.52 & --- & --- & -0.2 & 0.31/0.85:/0.54 & 2.07/1.39 & 1.16/1.47 & 1.12/1.49 & NA\\
CrA-432 & 19005974-3647109	&  M4.0 ($\pm$1.0)  &  2.7$\pm$0.2   &   -46 & --- & --- & --- & -2.8: & -0.34/-0.59/-0.25 & 2.43/1.89 & 1.12/1.50 & 1.18/1.53 & A\\
CrA-452 & 19004455-3702108	&  M2.5 ($\pm$0.5)  &  1.7$\pm$0.2  &   -2.6 & 0.51 & --- & --- & --- & 0.23:/0.83:/0.80 & 1.45/0.82 & 0.83/1.15 & 0.78/1.03 & NA\\
CrA-468$^*$ & 19014936-3700285	&   M3.5 ($\pm$0.5) &  0.6$\pm$0.7  &   -2.8 & 0.53 & --- & --- & --- & 0.5/0.7/0.7 & 1.77/1.07 & 0.99/1.28 & 1.02/1.26 & NA\\
G-1 & 19022708-3658132	&   M0.5 ($\pm$0.5) &  3.4$\pm$0.2   &  -3.8 & 0.58 & --- & --- & --- & 0.4/1.5/1.1 & 1.28/0.62 & 0.62/0.77 & 0.59/0.79 & A\\
G-32 & 19015833-3700267	&  ---   & ---   &   -4: & --- & --- & --- & --- & --- & --- & --- & --- & NA \\
G-48 & 19:01:49.8 -36:52:43	&  ---   &  ---   &  -9: & --- & --- & --- & --- & --- & --- & --- & --- & Uncertain \\
G-80 & 19:01:35.7 -37:02:32	&  ---   &  ---  &   -275: & --- & -25/-83 & -170:/-41: & --- & --- & --- & --- & --- & HH \\
G-85 & 19013385-3657448	&   M0.5 ($\pm$1.5) &  19$\pm$2  &   -31 & --- & --- & --- & --- & -2.6/-2.8/-2.1 & 2.27/1.06 & 0.72/0.98 & 0.76/0.98 & A \\
G-87 & 19013232-3658030	&   M1.5 ($\pm$1.5)  &  16$\pm$2  &   -3: & --- & --- & --- & --- & 0.15/0.52/0.91 & 2.16/2.79 & 0.51/0.89 & ---/1.56 & NA:\\
G-88 & 19:01:32.2 -36:51:20 	&  ---  &  ---  &   -13 & --- & --- & --- & --- & --- & --- & --- & --- & A:\\
G-90 & 19:01:31.4 -36:52:53 	&  ---  &  ---  &   -7.4 & --- & --- & --- & --- & --- & --- & --- & --- & A:\\
\enddata
\tablecomments{Lines observed in the FLAMES spectra. Uncertain values
are marked by ``:". The field ``ID/Coord." provides the
2MASS IDs of the objects, or their coordinates in case they were
not detected by 2MASS. The EW are given in \AA, with negative 
values used for emission lines. The wavelengths of the listed lines are
6562\AA\ (H$\alpha$), 6708\AA\ (Li I), 6548/6584\AA\ ([N II]),  6717/6731\AA\ ([S II]),
6678\AA\ (He I),8498/8542/8662\AA\ (Ca II IR triplet), 7665/7699\AA\ (K I), 8183/8185\AA\ (Na I).
For the Na I lines, we give two values measured
in the L773 and L881 spectra ('R' for red, 'I' for IR), respectively.
The comments list the presence of accretion (A) or no evidence for accretion
(NA), Herbig-Haro objects (HH). The accretion classification is made taking into account 
the H$\alpha$ EW, H$\alpha$ profile, and presence of the Ca II in emission.
$^*$ The object CrA-468 is also known as G-49 (SA08).}
\end{deluxetable}

\begin{deluxetable}{lcccccccl}
\tabletypesize{\scriptsize}
\rotate
\tablenum{2}
\tablecolumns{9} 
\tablewidth{0pc} 
\tablecaption{HAWK-I data for known Coronet members and potential members\label{hawki-table}} 
\tablehead{
 \colhead{Name}  & \colhead{RA (J2000)} & \colhead{DEC (J2000)} & \colhead{RA (deg)} & \colhead{DEC (deg)} & \colhead{J (mag)}   & \colhead{H (mag)} & \colhead{K (mag)} & \colhead{Comments}} 
\startdata
CrA-4107 & 19:02:54.64 & -36:46:19.1 &285.727661 & -36.771973 & 12.577$\pm$0.009 & 12.008$\pm$0.015 & 11.631$\pm$0.015 & variable (0.1-0.2 mag) \\
CrA-452 & 19:00:44.55 & -37:02:10.8 & 285.185638 & -37.036343 & 11.853$\pm$0.018 & --- & --- &  variable ($\sim$0.3 mag) \\
CrA-465 & 19:01:53.74 & -37:00:33.9 & 285.473907 & -37.009418 & 14.078$\pm$0.079 & 13.479$\pm$0.061 & 13.001$\pm$0.070 & \\
CrA-468 & 19:01:49.36 & -37:00:28.5 & 285.455658 & -37.007915 & 12.508$\pm$0.079 & 12.037$\pm$0.061 & --- &  also called G-49\\
G-14 & 19:02:12.01 & -37:03:09.3 & 285.550049 & -37.052582 & 13.453$\pm$0.052 & 12.633$\pm$0.029 & 12.200$\pm$0.027 &  \\
G-30 & 19:02:00.12 & -37:02:22.0 & 285.500488 & -37.039444 & 12.064$\pm$0.079 & --- & --- &  \\
G-32 & 19:01:58.33 & -37:00:26.7 & 285.493042 & -37.007416 & 18.134$\pm$0.080 & 15.418$\pm$0.061 & 13.650$\pm$0.070 &  \\
G-116 & 19:01:15.40 & -36:52:36.0 & 285.314178 & -36.876667 & 21.116$\pm$0.167 & 18.962$\pm$0.115 & 18.493$\pm$0.068 &  prob. extragal.  \\
G-119 & 19:01:07.90 & -36:51:21.0 & 285.282928 & -36.855831 & 19.252$\pm$0.039 & 18.347$\pm$0.038 & 17.831$\pm$0.040 &  prob. extragal. \\
\enddata
\tablecomments{HAWK-I JHK magnitudes for the detected known Coronet members
and potential members (based on Spitzer and X-ray data). G-116 and G-119 are probably extragalactic 
sources. }
\end{deluxetable}

\begin{deluxetable}{lccccl}
\tabletypesize{\scriptsize}
\tablenum{3}
\tablecolumns{6} 
\tablewidth{0pc} 
\tablecaption{Clumps and structures found in the LABOCA map \label{laboca-tab}} 
\tablehead{
 \colhead{Name} & \colhead{RA (J2000)} &  \colhead{DEC (J2000)}   & \colhead{Extension (beam)}& \colhead{Integrated Flux (Jy)}& \colhead{Related sources} } 
\startdata
1 & 19:01:53.9 & -36:58:03 & 96$\pm$8 & 33.3$\pm$0.8 & COA2, Ch10,11,12, see Fig. \ref{contours-fig} \\
2 & 19:01:42.5 & -36:58:39 & 6$\pm$1 & 1.44$\pm$0.02 & IRS2,Ch9 \\    
3 & 19:02:59.4 & -37:07:45 & 22$\pm$7 & 3.40$\pm$0.35 & Ch23, IRAS 18595-3712 \\  
4 & 19:03:07.9 & -37:13:04 & 10$\pm$2 & 1.65$\pm$0.07 & Ch24, VV CrA \\  
5 & 19:01:39.5 & -36:53:29 & 35$\pm$7 & 4.55$\pm$0.43 & HD-176386B,Ch6,7,8 \\    
6 & 19:02:18.3 & -37:01:41 & 27$\pm$20 & 2.1$\pm$1.1 & Ch21 \\    
7+10+24 & 19:03:06.4 & -37:14:53 & 32$\pm$5 & 2.40$\pm$0.15 & COA4 \\    
8+22 & 19:00:55.4 & -36:55:27 & 29$\pm$7 & 1.95$\pm$0.28 & \\     
9+14 & 19:01:57.0 & -37:01:05 & 20$\pm$5 & 1.43$\pm$0.18 & Ch14 \\    
11 & 19:02:11.4 & -36:56:22 & 15$\pm$7 & 1.08$\pm$0.33 & G-17 \\    
12 & 19:02:13.7 & -36:57:26 & 7$\pm$1 & 0.45$\pm$0.05 & Ch19 \\    
13+25 & 19:02:11.4 & -37:00:47 & 16$\pm$2 & 1.15$\pm$0.08 & Ch18: \\    
16 & 19:02:24.3 & -36:56:50 & 13$\pm$4 & 0.73$\pm$0.15 & \\     
17+20+26 & 19:01:05.3 & -36:54:14 & 35$\pm$13 & 1.88$\pm$0.53 & Ch2,3,4 \\    
18+21 & 19:01:13.7 & -36:54:23 & 27$\pm$6 & 1.33$\pm$0.03 & Ch5 \\    
19 & 19:02:33.5 & -37:02:08 & 13$\pm$6 & 0.63$\pm$0.03 & \\     
23 & 19:02:18.3 & -37:04:43 & 13$\pm$5 & 0.48$\pm$0.15 & \\     
27 & 19:01:09.1 & -36:57:26 & 6$\pm$2 & 0.77$\pm$0.02 & S-CrA,Ch1 \\    
\enddata
\tablecomments{Clumps detected with 2d Clumpfind in the 870$\mu$m map.
When the identification of a structure as one or several clumps depends on the parameters
used by 2d Clumpfind, we list it as a "multiple structure" (e.g. 7+10+24).
The last column indicates whether the clump is associated to known sources or
structures detected at different wavelengths. Millimeter clumps from Chini et al. (2003)
are marked as Ch\#. CO detections from Vilas-Boas et al. (2000) are marked as
COA\#. Uncertain values are marked with ``:". Note that the association to a source is merely
the coincidence of coordinates, it does not mean that the submillimeter clump and the
(optical, X-ray, CO) source are the same object (see text for details).}
\end{deluxetable}

\begin{deluxetable}{lcclll}
\tabletypesize{\scriptsize}
\rotate
\tablenum{4}
\tablecolumns{6} 
\tablewidth{0pc} 
\tablecaption{Spectral typing indices \label{index-table}} 
\tablehead{
 \colhead{Name} & \colhead{$\lambda$ Numerator} & \colhead{$\lambda$ Denominator} & \colhead{Range}   & \colhead{Calibration} & \colhead{Reference}} 
\startdata
PC1 & 7030-7050 & 7525-7550 &  M3-M9 & -0.06+2.95 X	 & 1,2 \\
PC2 & 7540-7580 & 7030-7050 &  M4-M8 & -0.63+3.89 X      & 1,2 \\
PC3 & 8235-8265 & 7540-7580 &  M3-M9 & -8.01+14.08 X-2.81 X$^2$ & 2 \\
PC4 & 9190-9225 & 7540-7580 &  M3-M9 & -0.94+4.66 X-0.52 X$^2$  & 2\\
R1  & 8025-8130 & 8015-8025 &  M2.5-M8 & 2.8078+21.085(X-1.044)-53.025(X-1.044)$^2$+60.755(X-1.044)$^3$ & 3 \\
R2  & 8415-8460 & 8460-8470 &  M3-M8 & 2.9091+10.503(X-1.035)-14.105(X-1.035)$^2$+8.5121(X-1.035)$^3$ & 3 \\
R3  & (8025-8130)+(8415-8460) & (8015-8025)+(8460-8470) & M2.5-M8 & 2.8379+19.708(X-1.035)-47.679(X-1.035)$^2$+52.531(X-1.035)$^3$ & 3 \\
TiO~8465 & 8405-8425 & 8455-8475 & M3-M8 & 3.2147+8.7311(X-1.085)-10.142(X-1.085)$^2$+5.6765(X-1.085)$^3$ & 3 \\
VO~2 & 7920-7960 & 8130-8150 & M3-M8 & 2.6102-7.9389(X-0.963)-8.3231(X-0.963)$^2$-14.660(X-0.963)$^3$ & 3 \\
VO~7445 & 0.5625(7350-7400)+0.4375(7510-7560) & 7420-7470 & M5-M8 & 5.0881+17.121(X-0.982)+13.078(X-0.982)$^2$ & 3 \\
\enddata
\tablecomments{Indices used for spectral typing. The wavelengths are given in \AA. 
In the calibration, X represents the index, and the resulting number indicates the
M subtype.
References: 1= Kirkpatrick et al. (1996); 2= Mart\'{\i}n et al. (1996); 3= Riddick et al. (2007).
The lineal relations for PC1 and PC2 are obtained by fitting together the objects in references
1 and 2.}
\end{deluxetable}

\begin{deluxetable}{lcccccl}
\tabletypesize{\scriptsize}
\rotate
\tablenum{5}
\tablecolumns{7} 
\tablewidth{0pc} 
\tablecaption{Fluxes for the individual sources detected with LABOCA\label{seds-table}} 
\tablehead{
\colhead{Wavelength ($\mu$m)/Filter} & \colhead{IRS2}  & \colhead{IRAS 18595-3712} & \colhead{VV CrA} & \colhead{HD 176386} & \colhead{S CrA} & \colhead{References}} 
\startdata
0.44/B	& ---	& ---	& ---			& 4.67		& 0.0335	& SIMBAD\\
0.55/V	& ---	& ---	& ---			& 4.38		& 0.0656	& SIMBAD\\
1.22/J	& 0.0050$\pm$0.0001	& 0.00007:	& 0.179$\pm$0.005	& 1.43$\pm$0.10	& 0.828$\pm$0.019 & 2MASS\\
1.63/H	& 0.129$\pm$0.004	& 0.0005	& 0.729$\pm$0.030	& 1.03$\pm$0.10	& 1.54$\pm$0.04	& 2MASS\\
2.19/K$_s$ & 0.949$\pm$0.023	& 0.0010$\pm$0.0001	& 2.028$\pm$0.040	& 0.89$\pm$0.08	& 2.29$\pm$0.05	& 2MASS\\
3.6	& ---	& 0.00976$\pm$0.00004	& 4.753$\pm$0.017	& 0.2059$\pm$0.0004	& 2.66$\pm$0.01	& IRAC/Spitzer\\
4.5	& ---	& 0.0224$\pm$0.0004	& 6.160$\pm$0.022	& 0.1388$\pm$0.0004	& 3.16$\pm$0.01	& IRAC/Spitzer\\
5.8	& ---	& 0.0224$\pm$0.0010	& 15.21$\pm$0.05	& $^a$		& 3.81$\pm$0.02	&IRAC/Spitzer\\
8.0	& 7.71$\pm$0.02	& 0.0138$\pm$0.0004	& ---			& $^a$		& 4.17$\pm$0.02	& IRAC/Spitzer\\
9.0	& ---	& ---	& 24.12$\pm$0.23	& ---	& 4.41$\pm$0.08	& AKARI\\
12	& 13:	& ---	& 31.9:	& ---	& ---	& IRAS\\
18	& ---	& 0.884$\pm$0.026	& 39.7$\pm$4.3		& ---	& 7.19$\pm$0.10	& AKARI\\
23.9	& ---	& 2.474$\pm$0.001	& ---	& ---	& ---	&\\
25	& 50:	& 3.69:	& 69.1:	& ---	& ---	& IRAS\\
60	& ---	& 38.9:	& 131:	& ---	& ---  & IRAS\\
65	& ---	& ---	& ---	& ---	& 13.8$\pm$4.3	& AKARI\\
100	& ---	& ---	& 95.2:	& ---	& ---  & IRAS\\
450	& 12$\pm$1	& ---	& ---	& ---	& 3$\pm$0.3	& Nutter et al.(2005)\\
850	& 2.0$\pm$0.7	& ---	& ---	& ---	& 0.7$\pm$0.2	& Nutter et al.(2005) \\
870	& 1.44$\pm$0.02	& 3.40$\pm$0.35	& 1.65$\pm$0.07	& 4.55$\pm$0.43$^b$	& 0.77$\pm$0.02	& This work\\
870	& ---	& ---	& ---	& ---	& 0.507$\pm$0.260	& Gezari et al. (1999)\\
1200	& 1.32$\pm$0.70	& 0.672$\pm$0.340	& ---	& ---	& 0.29$\pm$0.15	& Chini et al. (2003)	\\
\enddata
\tablecomments{Multiwavelength fluxes for the sources in Figure \ref{seds-fig}.
All the fluxes are in Jy.
The data include 2MASS JHK photometry and Spitzer data 
(Skrutskie et al. 2006; Currie \& Sicilia-Aguilar 2010), IRAS observations (Wilking et al. 1992),
AKARI data (9, 18, and 65$\mu$m; Kataza et al. 2010), and mm/submm points (Gezari et al. 1999; Chini et al. 2003;
Nutter et al. 2005). The Spitzer data for VV CrA and IRAS 18595-3712 has not been published
previously, but it was obtained as aperture photometry with MOPEX
on the BCD mosaics using the same techniques and parameters exposed in 
Currie \& Sicilia-Aguilar (2010). Uncertain values are marked by ``:".
$^a$ Extended emission, the source cannot be
extracted. $^b$ The emission is very extended and probably
includes part of the molecular cloud, in addition to the source.}
\end{deluxetable}

\begin{deluxetable}{lcccccc}
\tabletypesize{\scriptsize}
\rotate
\tablenum{6}
\tablecolumns{7} 
\tablewidth{0pc} 
\tablecaption{Summary of JHK data for sources in the HR diagram\label{jhk-table}} 
\tablehead{
 \colhead{Name}  & \colhead{ID} &  \colhead{J (mag)}   & \colhead{H (mag)} & \colhead{K (mag)} & \colhead{Sp. Type} & \colhead{A$_V$ (mag)} } 
\startdata
CrA-4108 & 19020968-3646345 & 12.580$\pm$0.023 & 11.972$\pm$0.023 & 11.634$\pm$0.019 & M3.5 & 1.2$\pm$ 1.1      \\
CrA-4110 & 19011629-3656282 & 12.954$\pm$0.021 & 12.314$\pm$0.021 & 11.897$\pm$0.021 & M5 & 1.3$\pm$ 0.4     \\
CrA-4111 & 19012083-3703027 & 13.233$\pm$0.024 & 12.687$\pm$0.023 & 12.402$\pm$0.024 & M4.5 & 0.1$\pm$ 0.5      \\
CrA-432 & 19005974-3647109 & 14.190$\pm$0.026 & 13.333$\pm$0.035 & 12.821$\pm$0.030 & M4.0 & 2.7$\pm$ 0.2      \\
CrA-452 & 19004455-3702108 & 11.571$\pm$0.024 & 10.766$\pm$0.025 & 10.447$\pm$0.023 & M2.5 & 1.7$\pm$ 0.2      \\
CrA-453 & 19010460-3701292 & 13.338$\pm$0.024 & 12.524$\pm$0.025 & 12.075$\pm$0.023 & M4.5 & 0.4$\pm$ 0.3      \\
CrA-465 & 19015374-3700339 & 14.084$\pm$0.026 & 13.401$\pm$0.033 & 13.015$\pm$0.033 & M7.5 & 0.2$\pm$ 1.0      \\
CrA-466 & 19011893-3658282 & 12.834$\pm$0.024 & 11.245$\pm$0.025 & 10.453$\pm$0.021 & M2.0 & 8.1$\pm$ 0.4      \\
CrA-468 & 19014936-3700285 & 12.498$\pm$0.026 & 11.891$\pm$0.025 & 11.603$\pm$0.024 & M3.5 & 0.6$\pm$ 0.7      \\
G-1 & 19022708-3658132 & 9.307$\pm$0.024 & 8.292$\pm$0.038 & 7.900$\pm$0.016 & M0.5 & 3.4$\pm$ 0.2      \\
G-14 & 19021201-3703093 & 13.408$\pm$0.029 & 12.582$\pm$0.029 & 12.145$\pm$0.021 & M4.5 & 1.9$\pm$ 0.1      \\
G-30 & 19020012-3702220 & 11.859$\pm$0.026 & 11.238$\pm$0.024 & 11.003$\pm$0.026 & M3.5 & 0.1$\pm$ 0.5      \\
G-65 & 19014041-3651422 & 13.898$\pm$0.032 & 11.629$\pm$0.023 & 10.481$\pm$0.019 & M1.5 & 13.8$\pm$ 0.5      \\
G-85 & 19013385-3657448 & 14.448$\pm$0.030 & 11.910$\pm$0.023 & 10.474$\pm$0.019 & M0.5 & 19$\pm$ 2      \\
G-87 & 19013232-3658030 & 14.853$\pm$0.037 & 12.622$\pm$0.029 & 11.432$\pm$0.023 & M1.5 & 16$\pm$ 2      \\
G-94 & 19012901-3701484 & 11.637$\pm$0.024 & 10.956$\pm$0.023 & 10.663$\pm$0.021 & M3.5 & 0.6$\pm$ 0.1      \\
G-95 & 19012872-3659317 & 10.828$\pm$0.023 &  9.591$\pm$0.025 &  9.002$\pm$0.019 & M1.0  & 5.0$\pm$0.4    \\
G-102 & 19012562-3704535 & 12.363$\pm$0.023 & 11.654$\pm$0.021 & 11.299$\pm$0.023 & M5.0 & 0.7$\pm$ 0.1      \\
\enddata
\tablecomments{Compilation of 2MASS magnitudes, spectral types, and extinctions of the members with
known spectral types used in Figure \ref{colormagiso-fig}. }
\end{deluxetable}

\clearpage

\clearpage

\clearpage

\clearpage

\clearpage

\clearpage

\clearpage

\clearpage

\clearpage

\clearpage


\begin{thebibliography}{}


\bibitem[Ballesteros-Paredes \& Hartmann(2007)]{ballesteros07} Ballesteros-Paredes, J., Hartmann, L., 2007, Rev. Mex. A.A., 43, 123

\bibitem[Baraffe et al.(1998)]{baraffe98} Baraffe, I.,Chabrier, G., Allard, F., \& Hauschild, P., 1998, AA, 337, 403

\bibitem[Bessell et al.(1998)]{bessell98} Bessell, M. S.; Castelli, F.; Plez, B., 1998, A\&A, 333, 231

\bibitem[Cardelli et al.(1989)]{car89} Cardelli, J.~A., Clayton,  G.~C., \& Mathis, J.~S.\ 1989, \apj, 345, 245

\bibitem[Chen et al.(1997)]{chen97} Chen, H.; Grenfell, T. G.; Myers, P. C.; Hughes, J. D., 1997, ApJ, 478, 295

\bibitem[Chini et al.(2003)]{chini03}Chini, R. et al. 2003, A\&A, 409, 235

\bibitem[Currie \& Sicilia-Aguilar(2011)]{currie11} Currie, Th., \& Sicilia-Aguilar, A., 2011, ApJ in press

\bibitem[De Muizon et al.(1980)]{demuizon80} De Muizon, M., Rouan, D., Lena, P., Nicollier, C., Wijnbergen, J., 1980, AA, 83, 140

\bibitem[Elmegreen(2000)]{2000ApJ...530..277E} Elmegreen, B.~G.\ 2000, \apj, 530, 277 

\bibitem[Fang et al.(2009)]{2009A&A...504..461F} Fang, M., van Boekel, R., Wang, W., Carmona, A., Sicilia-Aguilar, A., \& Henning, T.\ 2009, \aap, 504, 461 

\bibitem[Fedele et al.(2010)]{2010A&A...510A..72F} Fedele, D., van den Ancker, M.~E., Henning, T., Jayawardhana, R., \& Oliveira, J.~M.\ 2010, \aap, 510, A72 

\bibitem[Forbrich \& Preibisch(2007)]{forbrich07} Forbrich, J.; Preibisch, T., 2007, A\&A 475, 959

\bibitem[Garmire \& Garmire(2003)]{garmire03} Garmire, G., \& Garmire, A., 2003, Astron. Nachr. 324, 153 

\bibitem[Gezari et al.(1999)]{1999yCat.2225....0G} Gezari, D.~Y., Pitts, P.~S., \& Schmitz, M.\ 1999, VizieR Online Data Catalog, 2225, 0 

\bibitem[Groppi et al.(2004)]{groppi04} Groppi et al., 2004, ApJ, 612, 946;
 
\bibitem[Groppi et al.(2007)]{2007ApJ...670..489G} Groppi, C.~E., Hunter, T.~R., Blundell, R., \& Sandell, G.\ 2007, \apj, 670, 489 

\bibitem[Hamaguchi et al.(2005)]{hamaguchi05a} Hamaguchi, K., et al. 2005, ApJ 618, 360

\bibitem[Hartmann et al.(2001)]{hartmann01} Hartmann, L., Ballesteros-Paredes, J., Bergin, E., 2001, ApJ 562, 852 

\bibitem[Henning et al.(1994)]{henning94} Henning, Th.; Launhardt, R.; Steinacker, J.; Thamm, E., 1994, A\&A, 338, 223

\bibitem[Hern\'{a}ndez et al.(2007)]{hernandez07} Hern\'{a}ndez, J., Hartmann, L., Megeath, S.T.,et al., 2007, ApJ, 662, 1067

\bibitem[Kainulainen et al.(2009)]{kai09} Kainulainen, J., Beuther,  H., Henning, T., \& Plume, R.\ 2009, \aap, 508, L35

\bibitem[Kataza et al.(2010)]{kataza10}Kataza, H., Alfageme, C., Cassatella, A., Cox, N., Fujiwara, H., Ishihara, D., Oyabu, S., Salama, A., Takita, S., and Yamamura, I., 2010, AKARI-IRC Point Source Catalogue Release note Version 1.0

\bibitem[Kirkpatrick et al.(1995)]{kirkpatrick95} Kirkpatrick, J. D.; Henry, T. J.; Simons, D. A. 1995, AJ, 109, 797

\bibitem[Knude \& Hog(1998)]{knude98} Knude, J.; Hog, E. 1998, A\&A 338, 897

\bibitem[Lada et al.(1994)]{lad94} Lada, C.~J., Lada, E.~A., Clemens,  D.~P., \& Bally, J.\ 1994, \apj, 429, 694

\bibitem[Lawson et al.(2009)]{lawson09} Lawson, W. A.,Lyo, A.-R., Bessell, M.S., 2009, MNRAS, 400, L29

\bibitem[Lombardi \& Alves(2001)]{lom01} Lombardi, M., \& Alves, J.\  2001, \aap, 377, 1023

\bibitem[L\'{o}pez-Mart\'{\i} et al.(2005)]{lopezmarti05} L\'{o}pez-Mart\'{\i}, B.,  Eisl\"{o}ffel, J., Mundt, R., 2005, AA, 444, 175 

\bibitem[L{\'o}pez Mart{\'{\i}} et al.(2010)]{2010A&A...515A..31L} L{\'o}pez Mart{\'{\i}}, B.,  et al. \ 2010, \aap, 515, A31 

\bibitem[Loren(1979)]{loren79} Loren, R. B., 1979, ApJ, 227, 832

\bibitem[Marraco \& Rydgren(1981)]{marraco81}Marraco \& Rydgren 1981, AJ, 86, 62

\bibitem[Mart\'{\i}n et al.(1996)]{martin96} Mart\'{\i}n, E., Rebolo, R., Zapatero-Osorio, M.R., 1996, ApJ, 469, 706

\bibitem[Meyer \& Wilking(2009)]{2009PASP..121..350M} Meyer, M.~R., \& Wilking, B.~A.\ 2009, \pasp, 121, 350 

\bibitem[Muzerolle et al.(1998)]{muzerolle98} Muzerolle, J., Hartmann, L., Calvet, N., 1998, AJ 116, 455

\bibitem[Muzerolle et al.(2010)]{2010ApJ...708.1107M} Muzerolle, J., Allen, L.~E., Megeath, S.~T., Hern{\'a}ndez, J., \& Gutermuth, R.~A.\ 2010, \apj, 708, 1107 

\bibitem[Natta et al.(2005)]{natta05} Natta, A., Testi, L., Randich, S., Muzerolle, J., 2005, Mem. S.A.It., 76, 343

\bibitem[Neuh\"{a}user et al.(2000)]{neuhauser00} Neuh\"{a}user, R., et al., 2000, A\&A, 146, 323

\bibitem[Nisini et al.(2005)]{nisini05} Nisini, B.; Antoniucci, S.; Giannini, T.; Lorenzetti, D., 2005, A\&A, 429, 543

\bibitem[Nutter et al.(2005)]{2005MNRAS.357..975N} Nutter, D.~J., Ward-Thompson, D., \& Andr{\'e}, P.\ 2005, \mnras, 357, 975 

\bibitem[Ratzka et al.(2008)]{2008poii.conf..519R} Ratzka, T., Leinert, C., Przygodda, F., \& Wolf, S.\ 2008, The Power of Optical/IR Interferometry: Recent Scientific Results and 2nd Generation, 519 

\bibitem[Riddick et al.(2007)]{riddick07} Riddick, F. C.; Roche, P. F.; Lucas, P. W., 2007, MNRAS, 381, 1067

\bibitem[Sicilia-Aguilar et al.(2005))]{sicilia05} Sicilia-Aguilar, A.; Hartmann, L.; Hern\'{a}ndez, J.; Brice\~{n}o, C.; Calvet, N., 2005, AJ, 130, 188

\bibitem[Sicilia-Aguilar et al.(2006))]{sicilia06ir} Sicilia-Aguilar, A., et al., 2006,ApJ  638, 897

\bibitem[Sicilia-Aguilar et al.(2008))]{sicilia08} Sicilia-Aguilar, A.; Henning, Th.; Juh\'{a}sz, A.; Bouwman, J.; Garmire, G.; Garmire, A., 2008, ApJ, 687, 1145

\bibitem[Taylor \& Storey(1984)]{taylor04} Taylor \& Storey, 1984, MNRAS, 209, 5

\bibitem[Walter(1986)]{walter86} Walter, F., 1986, ApJ, 306, 573

\bibitem[Walter et al.(1997)]{walter97} Walter, F., et al., 1997, AJ, 114, 1544.

\bibitem[Wang et al. (2004)]{wang04} Wang, H., Mundt, R., Henning, Th., Apai, D., 2004, ApJ 617, 1191

\bibitem[White \& Basri(2003)]{white03} White, R., \& Basri, G., 2003, \apj, 582, 1109

\bibitem[Wilking et al.(1985)]{wilking85} Wilking et al., 1985, ApJ, 293, 165

\bibitem[Williams et al.(1994)]{williams94} Williams, J., et al. 1994, ApJ 428, 693

\end{thebibliography}
\end{document}